\documentclass[conference]{IEEEtran}
\IEEEoverridecommandlockouts

\usepackage{cite}
\usepackage{amsmath,amssymb,amsfonts}
\usepackage{algorithmic}
\usepackage{graphicx}
\usepackage{url}
\usepackage{booktabs}
\usepackage{array}
\usepackage{listings}
\usepackage{textcomp}
\usepackage{xcolor}
\usepackage{enumitem}
\usepackage[T1]{fontenc}
\usepackage{tikz}
\usetikzlibrary{positioning,fit,backgrounds,arrows.meta, calc}
\usepackage[hidelinks]{hyperref}
\usepackage{url}
\usepackage[acronym]{glossaries}
\usepackage{balance}
\usepackage{makecell}
\usepackage{colortbl}
\usepackage{booktabs}   
\usepackage{pifont}

\definecolor{codebg}{RGB}{245,245,245}
\definecolor{codecmt}{RGB}{110,110,110}
\definecolor{L1col}{HTML}{E87722}
\definecolor{L1bg}{HTML}{FFF3E6}
\definecolor{L2col}{HTML}{1565C0}
\definecolor{L2bg}{HTML}{E3F2FD}
\definecolor{L3col}{HTML}{2E7D32}
\definecolor{L3bg}{HTML}{E8F5E9}
\definecolor{L3Dcol}{HTML}{00796B}
\definecolor{L3Dbg}{HTML}{E0F2F1}
\definecolor{L4col}{HTML}{6A1B9A}
\definecolor{L4bg}{HTML}{F3E5F5}
\definecolor{L4Dcol}{HTML}{00838F}
\definecolor{L4Dbg}{HTML}{E0F7FA}

\usetikzlibrary{shapes.geometric, arrows, positioning}
\usepackage[most]{tcolorbox}
\tcbset{
    textmarker/.style={
        enhanced,
        parbox=false,
        boxrule=0mm,
        boxsep=0.5mm,
        arc=0mm,
        outer arc=0mm,
        left=1mm,
        right=0mm,
        top=0mm,
        bottom=0mm,
        toptitle=1mm,
        bottomtitle=1mm,
        oversize,
        frame hidden
    }
}
\newtcolorbox{futureworkBox}{
    textmarker,
    breakable,
    borderline west={1pt}{0pt}{green!60!black},
    colback=black!0!white,
}

\newacronym{RaaS}{RaaS}{Ransomware-as-a-Service}
\newacronym{ENTRAP}{ENTRAP}{Enterprise Network Testbed for Reproducible Attack and Defense Practices}
\newacronym{IaC}{IaC}{Infrastructure-as-Code}
\newacronym{DMZ}{DMZ}{Demilitarized Zone}
\newacronym{C2}{C2}{Command-and-Control}
\newacronym{MITM}{MITM}{Man-in-the-Middle}
\newacronym{WAN}{WAN}{Wide Area Network}
\newacronym{NAT}{NAT}{Network Address Translation}
\newacronym{DNAT}{DNAT}{Destination NAT}
\newacronym{SDN}{SDN}{Software-Defined Networking}
\newacronym{IDS}{IDS}{Intrusion Detection System}
\newacronym{IPS}{IPS}{Intrusion Prevention System}
\newacronym{ML}{ML}{Machine Learning}
\newacronym{DL}{DL}{Deep Learning}
\newacronym{VM}{VM}{Virtual Machine}

\def\BibTeX{{\rm B\kern-.05em{\sc i\kern-.025em b}\kern-.08em
    T\kern-.1667em\lower.7ex\hbox{E}\kern-.125emX}}
\begin{document}

\title{Experience: ENTRAP -- An IaC Enterprise Cyber Range for Reproducible Security Assessment and Dataset Generation}

\author{\IEEEauthorblockN{Rathan Appana, Aditya Kulkarni, and Bheemarjuna Reddy Tamma}
\IEEEauthorblockA{\emph{Center for Cryptography and Cybersecurity, Indian Institute of Technology Hyderabad, India} \\
\{rathan.appana, aditya.kulkarni\}@prjt.cse.iith.ac.in, tbr@iith.ac.in}
}
\maketitle

\begin{abstract}
An enterprise breach causes most of its damage after the perimeter is crossed. Once an attacker gains a foothold, they can move laterally to other hosts, take over the trusted services of the network, and encrypt the backups needed for recovery. Defenders, on the other hand, need to study this post-compromise stage, but they cannot rehearse it on a production enterprise network. Building such an enterprise network manually is time-consuming, and resetting it to its clean state between runs is still harder. As a result, security experiments are difficult to perform, compare and reproduce.

We develop \gls{ENTRAP}, an \gls{IaC} cyber range that defines a fully segmented enterprise as version-controlled automation. ENTRAP uses Ansible, KVM/Libvirt, and Cloud-init to provision the environment and reset it to a clean state whenever necessary. The enterprise network includes several entities such as a gateway, an internal \gls{DMZ}, DNS, DHCP, a backup server, a web server, and clients. On this network, a user can learn to assemble the entities, launch attacks, and write rules or defensive measures that harden its security. We launch three attacks: a \gls{RaaS}-backed ransomware campaign, a Layer-2 DHCP starvation and rogue DHCP attack, and perimeter web exploitation. Each attack runs from a clean state, and ENTRAP acts as a reproducible generator of labeled attack datasets. Using the collected dataset, we train a Random Forest classifier for ransomware detection and present its evaluation results. We publicly release the codebase, collected logs, and datasets to facilitate further research.
\end{abstract}

\begin{IEEEkeywords}
Cyber Range, Infrastructure-as-Code, Intrusion Detection, Ransomware, Datasets.
\end{IEEEkeywords}

\glsresetall

\section{Introduction}
\label{sec:Introduction}
Enterprise networks form the backbone of modern organizations by connecting users, services, and information across different organizational segments, all governed by security policies. With a massive amount of sensitive data flowing through these networks, they become the primary target for cyberattacks. As organizations grow to include several more entities, they become increasingly dependent on these networks to process routine operations; thereby, the surface area for potential compromise grows proportionally. Despite years of investment and improvements in security tooling and awareness, attackers have kept pace by continuing the trajectory of the overall cost of data breaches. IBM's 2025 Cost of a Data Breach report~\cite{ibm2025costofdatabreach} recorded a global average breach cost of USD~4.4~million, down by 9\% from the previous year. This first decline in five years was driven largely by faster breach containment through AI-powered defenses. However, this improvement did not last longer; IBM's 2026 report~\cite{ibm2026costofdatabreach} recorded a global average cost of USD~4.99~million, a 12\% increase and an all-time high driven by a 56\% rise in AI-powered attacks. Additionally, according to the Verizon 2025 report~\cite{verizon2025dbir}, ransomware remains a primary driver of cyberattacks, present in 44\% of breaches. 

\subsection{Cyber Ranges and Automated Enterprise Network Testbed}
Studying the attack surface of an enterprise network requires a controlled 
experimental environment that reproduces its segments, services, and
inter-entity trust relationships, such as a client trusting its DHCP for network configurations. 
In such an environment, researchers can launch multi-layer attacks (network and application), collect labeled datasets, experiment with defensive strategies, and train personnel, all without impacting the production network.
A cyber range provides this by offering purpose-built network environments to safely launch cyberattacks, study their behavior across individual entities and the network as a whole, and easily reset between experiments. Such an environment allows analysis of the complete attack lifecycle in isolation.
However, a cyber range requires an underlying testbed: an emulated enterprise network that reproduces the organization's topology, services, and security policies. Manually emulating such a network requires significant time and effort and, more importantly, it cannot be easily reset to a clean state between experiments. This motivates the need for an \textit{automated} enterprise network testbed, one that 
can be described as code, instantiated on demand, torn down, and rebuilt identically any number of times. In this work, we present \gls{ENTRAP}: an automated testbed that instantiates the cyber range for enterprise environments.

\subsection{Motivation and Research Gaps}
Reproducible cyber ranges are not new, but existing ones leave certain gaps that we target in this work. Many published cyber ranges~\cite{yamin2020cyber,pham2016cyris,beuran2017cytrone} are oriented toward the human-facing training part rather than toward a reproducible offense-and-defense experimentation part. The cyber ranges that automate environment creation tend to target either per-trainee content~\cite{pham2016cyris,beuran2017cytrone}, or specialized tactical networks~\cite{chadha2016cybervan} rather than the segmented enterprise environments, which includes gateway, \gls{DMZ}, infrastructure services,  and client systems. Our work, \gls{ENTRAP}, is designed to fill this gap and serves as a fully automated enterprise cyber range that allows cybersecurity researchers to launch multi-layer attacks against a segmented enterprise network, collect labeled datasets from each experiment, and study the effectiveness of various defense mechanisms in a controlled environment.

\subsection{Our Contributions}
The key contributions of our work are as follows.
\begin{enumerate}[label=(\roman*)]
    \item We develop \gls{ENTRAP}, a fully automated, \gls{IaC} enterprise testbed that instantiates a complete segmented enterprise network reproducibly from a clean state on demand.
    \item On this testbed, we launch three attacks, (i)~\gls{RaaS}-backed ransomware campaign, (ii)~a Layer-2 DHCP starvation combined with rogue DHCP server, and (iii)~perimeter web exploitation, and we draft defensive mechanisms for each attack.
    \item We train a ransomware detection model on the collected dataset and evaluate its robustness. We release the codebase of ENTRAP, collected logs, and datasets publicly for further research~\cite{ENTRAP_GitHub}. 
\end{enumerate}


\section{Related Work}
\label{sec:Related_Work}
This section describes various cyber ranges, attacks performed on enterprise networks, and network defenses to mitigate such attacks.

\subsection{Reproducible Testbeds and Cyber Ranges}
For years, network experimentation has faced a major hurdle: \textit{reproducibility}. Table~\ref{tab:ENTRAP_comparison} compares ENTRAP with existing cyber ranges across these dimensions. Emulab~\cite{white2002integrated} filled this gap by introducing an experimental network configuration that, once set, can be instantiated whenever required. It unified emulated, simulated, and wide-area resources under a shared \texttt{ns}-language script. When scaling these environments became the next challenge, subsequent work~\cite{hibler2008large} multiplexed virtual hosts, routers, and links across fewer physical machines, which enabled them to run topologies an order of magnitude larger.

Cyber ranges take reproducible testbeds a step further by focusing specifically on security experiments. Yamin \textit{et al.}~\cite{yamin2020cyber} surveyed and evaluated these platforms across dimensions such as scenario generation, monitoring, teaming, scoring, and management, noting that many cyber range platforms heavily prioritize security training and human-in-the-loop exercises. To address the need of automation, CyRIS~\cite{pham2016cyris} launches cyber ranges on KVM from \texttt{YAML} descriptions. It populates these environments with vital security elements, including malware, attacks, captured traffic, and logs. 
Its generated ranges remain limited, usually confined to a single bus topology. CyTrONE~\cite{beuran2017cytrone} uses CyRIS as its backend engine to instantiate the environment and adds the training layer on top, including an instructor wizard that builds the exercise content and delivers it through an online learning platform, scaling smoothly to hundreds of trainees at the same time. 
With a different architectural approach, CyberVAN~\cite{chadha2016cybervan} provides a high-fidelity, GUI-based testbed that combines physical machines, virtual machines and network simulation to reproduce tactical and strategic military networks. However, its focus is high-fidelity military network scenarios operated as a closed-laboratory facility, rather than a segmented enterprise network expressed as reproducible, version-controlled infrastructure. DETERLab~\cite{DETERLab_Project} is a shared, Emulab-based cybersecurity experimentation infrastructure that supports physical, virtual, and containerized resources for large-scale experiments.

\subsection{Enterprise Attack and Network Defenses}
Normally, enterprise network security experiments focus on behavior after the initial compromise. MITRE ATT\&CK~\cite{mitreattack} organizes this behavior into several stages including initial access, discovery, lateral movement, and impact. Kharraz \textit{et al.}~\cite{kharraz2015cutting} show that the consequences of attacks such as ransomware strongly depend on how far an attacker can propagate within an internal network. These observations motivate test environments that not only represent the initial compromise but also the internal services, credentials, protocols, and trust relationships that enable movement between entities.

\begin{table}[t]
    \centering
    \caption{Comparison of ENTRAP with existing cyber ranges}
    \label{tab:ENTRAP_comparison}
    \resizebox{0.99\columnwidth}{!}{
    \begin{tabular}{@{}p{1.5cm} p{4.2cm} p{3.5cm}@{}}
        \toprule
        \textbf{Cyber range} & \textbf{Target environment} & \textbf{How the environment is defined?} \\
        \midrule
        Emulab~\cite{white2002integrated} & Generic network topologies; not security-specific & \texttt{ns} script \\
        CyRIS~\cite{pham2016cyris} & Single-host training ranges on a bus topology & YAML description; per-trainee instantiation \\
        CyTrONE~\cite{beuran2017cytrone} & Per-trainee training exercise (using CyRIS) & YAML description \\
        CyberVAN~\cite{chadha2016cybervan} & Tactical/Strategic military network scenarios & XML scenario file; GUI-based \\
        DETERLab~\cite{DETERLab_Project}& Shared infrastructure for large-scale cybersecurity experiments & Emulab-based testbed; experiment definitions and MAGI workflows \\
        \midrule
        \textbf{\gls{ENTRAP}} & Segmented enterprise network with \gls{DMZ} and inline IDS/IPS & IaC (Ansible, Cloud-init); clean state rebuild per run \\
        \bottomrule
    \end{tabular}}
\end{table}

Evaluating defense against infrastructure-level threats requires replicating inherently vulnerable protocols in a controlled environment. The unauthenticated broadcast design of DHCP in RFC~2131 exposes enterprise segments to address pool starvation and rogue responses; although a cryptographic message authentication exists in RFC~3118, its deployment rate remains very low. Instead, DHCP snooping at the switch level blocks rogue server responses from untrusted ports and is the standard approach, but software bridge environments typically lack this feature. Validating enterprise resilience requires observing how inline IDS/IPS systems such as Snort and Suricata react to these infrastructure exploits.

Existing platforms target training exercises (CyRIS, CyTrONE), tactical networks (CyberVAN), or shared cybersecurity experimentation at scale (DETERLab). However, these platforms do not provide a locally deployable, IaC-based enterprise network testbed that combines segmented topology, vulnerable infrastructure services, and inline IDS/IPS within an environment reconstructible from a known clean state.
\section{ENTRAP: Proposed Testbed}
\label{sec:ENTRAP_Proposed_Testbed}

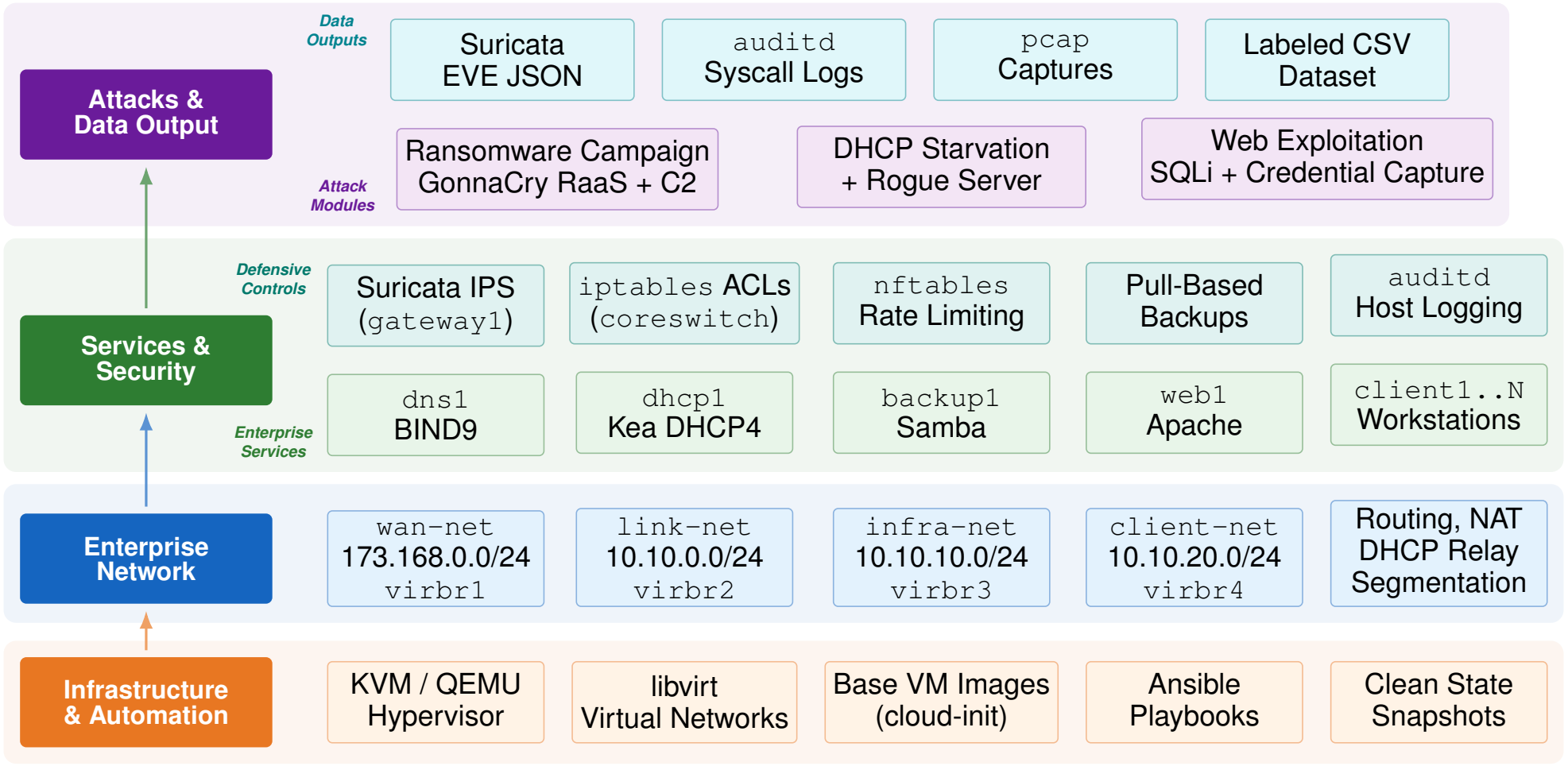
\begin{figure*}[t]
    \centering
    \scalebox{0.75}{
    \begin{tikzpicture}[
        font=\sffamily\small,
        sA/.style={
            draw=L1col!50,
            fill=L1bg,
            rounded corners=1.5pt,
            minimum height=9mm,
            minimum width=24mm,
            align=center,
            inner sep=2.5pt
        },
        sB/.style={
            draw=L2col!50,
            fill=L2bg,
            rounded corners=1.5pt,
            minimum height=9mm,
            minimum width=24mm,
            align=center,
            inner sep=2.5pt
        },
        sC/.style={
            draw=L3col!50,
            fill=L3bg,
            rounded corners=1.5pt,
            minimum height=9mm,
            minimum width=24mm,
            align=center,
            inner sep=2.5pt
        },
        sD/.style={
            draw=L3Dcol!50,
            fill=L3Dbg,
            rounded corners=1.5pt,
            minimum height=9mm,
            minimum width=24mm,
            align=center,
            inner sep=2.5pt
        },
        sE/.style={
            draw=L4col!50,
            fill=L4bg,
            rounded corners=1.5pt,
            minimum height=9mm,
            minimum width=32mm,
            align=center,
            inner sep=2.5pt
        },
        sF/.style={
            draw=L4Dcol!50,
            fill=L4Dbg,
            rounded corners=1.5pt,
            minimum height=9mm,
            minimum width=27mm,
            align=center,
            inner sep=2.5pt
        },
        tag/.style={
            font=\sffamily\footnotesize\bfseries,
            text=white,
            fill=#1,
            minimum height=10mm,
            minimum width=28mm,
            align=center,
            rounded corners=2pt,
            inner sep=3pt
        },
        rlbl/.style={
            font=\sffamily\tiny\bfseries\itshape,
            text=#1,
            align=center
        },
        band/.style={
            fill=#1,
            fill opacity=0.07,
            rounded corners=4pt,
            inner sep=5pt
        }
    ]
 
    \node[sA] (A1) {KVM / QEMU\\Hypervisor};
    \node[sA, right=3mm of A1] (A2) {libvirt\\Virtual Networks};
    \node[sA, right=3mm of A2] (A3) {Base VM Images\\(cloud-init)};
    \node[sA, right=3mm of A3] (A4) {Ansible\\Playbooks};
    \node[sA, right=3mm of A4] (A5) {Clean State\\Snapshots};
    \node[tag=L1col, left=6mm of A1] (T1) {Infrastructure\\[-1pt]\& Automation};
 
    \node[sB, above=6mm of A1] (B1) {\texttt{wan-net}\\173.168.0.0/24\\\texttt{virbr1}};
    \node[sB, above=6mm of A2] (B2) {\texttt{link-net}\\10.10.0.0/24\\\texttt{virbr2}};
    \node[sB, above=6mm of A3] (B3) {\texttt{infra-net}\\10.10.10.0/24\\\texttt{virbr3}};
    \node[sB, above=6mm of A4] (B4) {\texttt{client-net}\\10.10.20.0/24\\\texttt{virbr4}};
    \node[sB, above=6mm of A5] (B5) {Routing, NAT\\DHCP Relay\\Segmentation};
    \node[tag=L2col, left=6mm of B1] (T2) {Enterprise\\[-1pt]Network};
 
    \node[sC, above=6mm of B1] (C1) {\texttt{dns1}\\BIND9};
    \node[sC, above=6mm of B2] (C2) {\texttt{dhcp1}\\Kea DHCP4};
    \node[sC, above=6mm of B3] (C3) {\texttt{backup1}\\Samba};
    \node[sC, above=6mm of B4] (C4) {\texttt{web1}\\Apache};
    \node[sC, above=6mm of B5] (C5) {\texttt{client1..N}\\Workstations};
    \node[rlbl=L3col, left=6mm of C1, anchor=north] {Enterprise\\Services};
 
    \node[sD, above=3mm of C1] (D1) {Suricata IPS\\(\texttt{gateway1})};
    \node[sD, above=3mm of C2] (D2) {\texttt{iptables} ACLs\\(\texttt{coreswitch})};
    \node[sD, above=3mm of C3] (D3) {\texttt{nftables}\\Rate Limiting};
    \node[sD, above=3mm of C4] (D4) {Pull-Based\\Backups};
    \node[sD, above=3mm of C5] (D5) {\texttt{auditd}\\Host Logging};
    \node[rlbl=L3Dcol, left=6mm of D1, anchor=south] {Defensive\\Controls};
 
    \path (C1.center) -- (D1.center) coordinate[midway] (L3mid);
    \node[tag=L3col] (T3) at (T2.center |- L3mid) {Services \&\\[-1pt]Security};
 
    \node[sE, above=6mm of D1, xshift=13.5mm] (E1) {Ransomware Campaign\\GonnaCry RaaS + C2};
    \node[sE, above=6mm of D3] (E2) {DHCP Starvation\\+ Rogue Server};
    \node[sE, above=6mm of D5, xshift=-13.5mm] (E3) {Web Exploitation\\SQLi + Credential Capture};
    \node[rlbl=L4col, left=6mm of E1, anchor=north] {Attack\\Modules};
 
    \node[sF, above=3mm of E1, xshift=-5mm] (F1) {Suricata\\EVE JSON};
    \node[sF, right=3mm of F1] (F2) {\texttt{auditd}\\Syscall Logs};
    \node[sF, right=3mm of F2] (F3) {\texttt{pcap}\\Captures};
    \node[sF, right=3mm of F3] (F4) {Labeled CSV\\Dataset};
    \node[rlbl=L4Dcol, left=6mm of F1, anchor=south] {Data\\Outputs};
 
    \path (E1.center) -- (F1.center) coordinate[midway] (L4mid);
    \node[tag=L4col] (T4) at (T3.center |- L4mid) {Attacks \&\\[-1pt]Data Output};
 
    \begin{scope}[on background layer]
        \node[band=L1col, fit=(T1)(A1)(A5)] {};
        \node[band=L2col, fit=(T2)(B1)(B5)] {};
        \node[band=L3col, fit=(T3)(C1)(C5)(D1)(D5)] {};
        \node[band=L4col, fit=(T4)(E1)(E3)(F1)(F4)] {};
    \end{scope}
 
    \draw[-{Latex[length=2mm]}, thick, L1col!70] ([yshift=2pt]T1.north) -- ([yshift=-2pt]T2.south);
    \draw[-{Latex[length=2mm]}, thick, L2col!70] ([yshift=2pt]T2.north) -- ([yshift=-2pt]T3.south);
    \draw[-{Latex[length=2mm]}, thick, L3col!70] ([yshift=2pt]T3.north) -- ([yshift=-2pt]T4.south);
 
    \end{tikzpicture}}
    \caption{ENTRAP architectural overview (read bottom-up). Infrastructure automation, enterprise network segments, services paired with defensive controls, and attack modules with their resulting outputs.}
    \label{fig:ENTRAP_Block_Diagram}
\end{figure*}
\gls{ENTRAP} is an enterprise cyber range expressed entirely as \gls{IaC}, reproducing the segments of a real enterprise network. Figure~\ref{fig:ENTRAP_Block_Diagram} outlines its four layers (read bottom-up): (i)~\textit{infrastructure and automation layer}, which provisions all entities using KVM/QEMU, \texttt{libvirt}, and Ansible, (ii)~\textit{enterprise network layer}, which defines four isolated segments, (iii)~\textit{services and security layer}, which details the services and defensive controls built in each entity, and (iv)~\textit{attacks and data output layer}, which illustrates the attack launched on testbed and the logs recorded during each attack.

The goal of the design is reproducibility: every run starts from the same clean state, so an experiment (offensive) that damages or breaches the environment can be reset before the next run. 
A clean state refers to a freshly provisioned deployment in which all VMs, virtual networks, service configurations, and experiment-generated state from the previous run have been removed and recreated using the same ENTRAP automation.
It also provides the feasibility of a working attack toolkit and an inline detection stack; as a result, both offensive and defensive mechanisms can be exercised repeatedly and reset between runs.


\subsection{Technology Stack and Software Used}
We use KVM/QEMU with \texttt{libvirt}, which provides the virtualization layer on which every entity runs as a \gls{VM}. Cloud-init provisions each VM at first boot by setting its hostname, network interfaces, users, and specified packages. Ansible automates the entire setup: it defines the entities, applies roles to them, and configures services in a fixed order. All entities share the Ubuntu~22.04 server cloud image as a common base. Network enforcement and services use \texttt{nftables} on the gateway, \texttt{iptables} on the internal router, Kea DHCP4 with an ISC DHCP relay on the router, BIND9 for the DNS server, Samba with \texttt{rsync} for the backup server, and Apache2 with PHP for the web server. Suricata~\cite{suricata} runs inline and uses a set of local and community rules to protect the enterprise network.
None of these entities is novel; they are designed on established standards. The key contribution is that ENTRAP assembles these entities in a fixed order and re-instantiates them identically on demand: seven entities across four isolated segments.
\begin{table}[t]
    \centering
    \caption{Virtual machines and their roles}
    \label{tab:vms}
    \scriptsize
    \setlength{\tabcolsep}{3pt}
    \begin{tabular}{@{}lll@{}}
    \toprule
    \textbf{VM} & \textbf{IP address(es)} & \textbf{Role \& key technology} \\
    \midrule
    \texttt{gateway1}
        & \makecell[l]{10.10.0.1\,(LAN)\\ \quad173.168.0.10\,(WAN)}
        & \makecell[l]{NAT edge router; Suricata IPS inline via \\ \texttt{nftables}/\texttt{NFQUEUE}; DNATs :80\,$\rightarrow$\,\texttt{web1}} \\[2pt]

    \texttt{coreswitch}
        & \makecell[l]{10.10.0.2\,(uplink)\\\quad10.10.10.1\,(infra)\\\quad10.10.20.1\,(client)}
        & \makecell[l]{Internal router / DMZ boundary;\\stateful \texttt{iptables}; \texttt{isc-dhcp-relay}} \\[2pt]

    \texttt{dns1}
        & 10.10.10.53
        & BIND9; authoritative for \texttt{enterprise.lab} \\[2pt]

    \texttt{dhcp1}
        & 10.10.10.2
        & Kea DHCP4; pools .150\,--\,.200 per segment \\[2pt]

    \texttt{backup1}
        & 10.10.10.20
        & Samba + \texttt{rsync}; nightly SSH-pull backup \\[2pt]

    \texttt{web1}
        & 10.10.10.10
        & Apache2/PHP; deliberately vulnerable \\[2pt]

    \texttt{client1..N}
        & 10.10.20.150\,--\,.200
        & Ubuntu workstations; DHCP-assigned \\[2pt]

    \texttt{c2server}
        & 173.168.0.50
        & Flask RaaS C2; GonnaCry payload server \\
    \bottomrule
    \end{tabular}
\end{table}

\subsection{VMs and their Roles}
Table~\ref{tab:vms} lists every entity in the testbed with IP addresses and key technologies used. The gateway (\texttt{gateway1}) is the only entity that performs \gls{NAT}. The internal router (\texttt{coreswitch}) routes traffic between the infrastructure and client segments and enforces access control policy at the segment boundary; although \texttt{web1} shares the infrastructure subnet with the other services, coreswitch ACLs restrict inter-segment traffic, creating a logical \gls{DMZ} boundary without a physically separate network segment. Its name reflects the topological position as the core of the network, although it operates at Layer~3. The remaining internal entities serve as the DNS, DHCP, backup server, and web server, alongside client workstations. Finally, a dedicated attacker-controlled entity, \texttt{c2server}, sits on the simulated \gls{WAN}, and it is referred to during the ransomware campaign described in Section~\ref{sec:Offensive_Measures}.

\subsection{Virtual Networks}
The enterprise network is divided into four isolated \texttt{libvirt} networks (see Figure~\ref{fig:ENTRAP_Block_Diagram}): a simulated WAN (\texttt{wan-net}), a gateway-to-core transit link (\texttt{link-net}), an infrastructure segment (\texttt{infra-net}), and a client segment (\texttt{client-net}). Each network is completely separated into its own virtual switch (\texttt{virbr1}--\texttt{virbr4}) and does not allow cross bridge connectivity at Layer~2. The gateway is the only entity that masquerades outbound traffic to the simulated WAN, and the only inline path from WAN is a \gls{DNAT} rule that exposes \texttt{web1} on TCP:80. Ansible must run on the hypervisor because the internal networks have no outside access. Ansible instead uses an SSH ProxyJump via \texttt{gateway1} to reach each of them.

\subsection{Network Topology}
\begin{figure}[t]
    \centering
    \scalebox{0.88}{
    \begin{tikzpicture}[
        font=\scriptsize,
        node distance=5mm,
        box/.style={
            draw,
            rounded corners=2pt,
            align=center,
            inner sep=3pt,
            minimum height=7mm,
            fill=white
        },
        net/.style={
            draw,
            dashed,
            rounded corners=3pt,
            inner sep=5pt
        },
        link/.style={
            -{Latex[length=1.6mm]},
            thick
        }
    ]
        \node[box] (wan) {Simulated WAN\\\texttt{wan-net} 173.168.0.0/24};
        \node[box, below=3mm of wan] (gw) {\texttt{gateway1}\\NAT + \texttt{nftables} + Suricata IPS};
        \node[box, below=3mm of gw] (link) {\texttt{link-net} 10.10.0.0/24};
        \node[box, below=3mm of link] (core) {\texttt{coreswitch}\\\texttt{iptables} DMZ ACL + DHCP relay};
        \node[box, below left=3mm and -1.5cm of core] (infra) {\texttt{infra-net} 10.10.10.0/24\\\texttt{dns1}, \texttt{dhcp1}, \texttt{backup1}, \texttt{web1}};
        \node[box, below right=3mm and -1.5cm of core] (client) {\texttt{client-net} 10.10.20.0/24\\\texttt{client1..N}};
        \draw[link] (wan) -- (gw);
        \draw[link] (gw) -- (link);
        \draw[link] (link) -- (core);
        \draw[link] (core) -- (infra);
        \draw[link] (core) -- (client);
    \end{tikzpicture}}
    \caption{Network topology of the ENTRAP cyber range.}
    \label{fig:topology}
\end{figure}
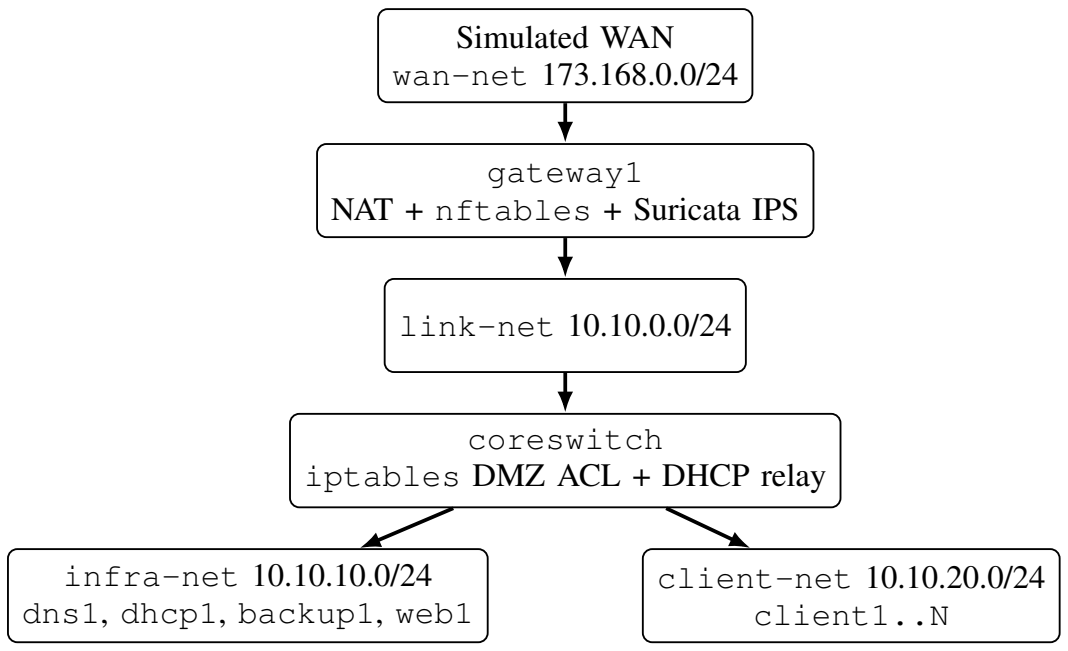

Figure~\ref{fig:topology} presents the network topology of our enterprise network testbed, which consists of several entities. The gateway separates the simulated WAN from the enterprise and runs the inline IDS/IPS. The internal router sits between the infrastructure and client segments and is the single point of DMZ access control policy. All deep packet inspection occurs at the gateway, while the internal router applies rules specific to address- and port-levels.

\section{Offensive and Defensive Measures}
\label{sec:Offensive_and_Defensive_Measures}
This section states the attacker model, launching three independent attacks on the enterprise network: a \gls{RaaS} campaign, a Layer-2 DHCP starvation with a rogue DHCP server, and a perimeter web exploitation, and describes the defensive mechanisms to detect and mitigate each of these attacks. We also tabulate the testbed instantiation time from a clean state and summarize the network flows recorded at the gateway.

\subsection{Attacker Model}
\label{sec:Attacker_Model}
\noindent\textbf{Attacker Identity:} We consider a single attacker, \textit{Trudy}, whose starting position and privileges vary per attack.
\begin{itemize}[nosep, leftmargin=*]
    \item \textit{Ransomware campaign:} Trudy is an \textit{external} attacker who controls \texttt{c2server} (173.168.0.50) on the simulated \gls{WAN}. Trudy has already performed an initial compromise phase (e.g., a phishing click or a leaked weak password) on a client. This is an assumed precondition, and not a demonstrated exploit.
    \item \textit{DHCP starvation and rogue server:} Trudy is an \textit{insider} on the client segment (\texttt{client-net}, 10.10.20.0/24), and operates from \texttt{client2}.
    \item \textit{Perimeter web exploitation:} Trudy is an \textit{external} attacker operating from the simulated \gls{WAN} (\texttt{wan-net}, 173.168.0.0/24) without having an internal access.
\end{itemize}

\noindent\textbf{In scope:} Across all three attacks performed, Trudy can send and receive arbitrary traffic, craft Layer-2 frames, spoof MAC addresses, and run standard tools (\texttt{Scapy}, \texttt{dnsmasq}, \texttt{nmap}). Trudy can also reuse any keys recovered on one entity against every other entity. Trudy, with these three capabilities, can exercise the three attacks on the enterprise network: (i)~delivery and execution of ransomware attack on a compromised client, lateral movement, and encryption, (ii)~exhausting the IP address in the legitimate DHCP pool, and setting up a rogue server with configurations on gateway IP address and DNS resolver address, (iii)~exploitation of a deliberately vulnerable web server.

\noindent\textbf{Out of scope:} Trudy does not control the hypervisor, physical bridges, or the gateway's firewall rules. She also cannot access the host machine, compromise the base VM image, or tamper with any of the Ansible playbooks that setup the very enterprise network.

\noindent\textbf{Role of the C2 server:} The \texttt{c2server} (173.168.0.50) is only used for the ransomware campaign. It hosts the GonnaCry \gls{RaaS}, which is a web program (built with Flask) that does three things: it hands out the encryptor and decryptor binaries, tracks each victim with a unique machine ID, and manages the key-exchange protocol for decryption. This server has no role to play in the DHCP attack and the web exploitation attack.

Note that the communications between the compromised client and the C2 server uses plaintext HTTP protocol on port~80. ENTRAP does not include a PKI or certification authority, so a trusted TLS certificate cannot be issued within it. Also, this is a deliberate design choice we make so that the Suricata rules inline at the \texttt{gateway1} can perform deep packet inspection and check if the traffic matches with the rules. We acknowledge that even under HTTPS with TLS~1.3, the SNI field in the \texttt{ClientHello} message is in plaintext and Suricata can still identify if the packet is directed to the C2 server. However, we do not implement SNI-based detection in the current version of the ENTRAP testbed. Also, Encrypted Client Hello (ECH) conceals the SNI field as well, but we have left these two approaches as a part of our future work.

\noindent\textbf{Correlation of attacks:} The three attacks are run independently each starting from a clean state of the enterprise network. No attack depends on the outcome of the other.

\subsection{Offensive Measures}
\label{sec:Offensive_Measures}
\subsubsection{Ransomware Campaign}
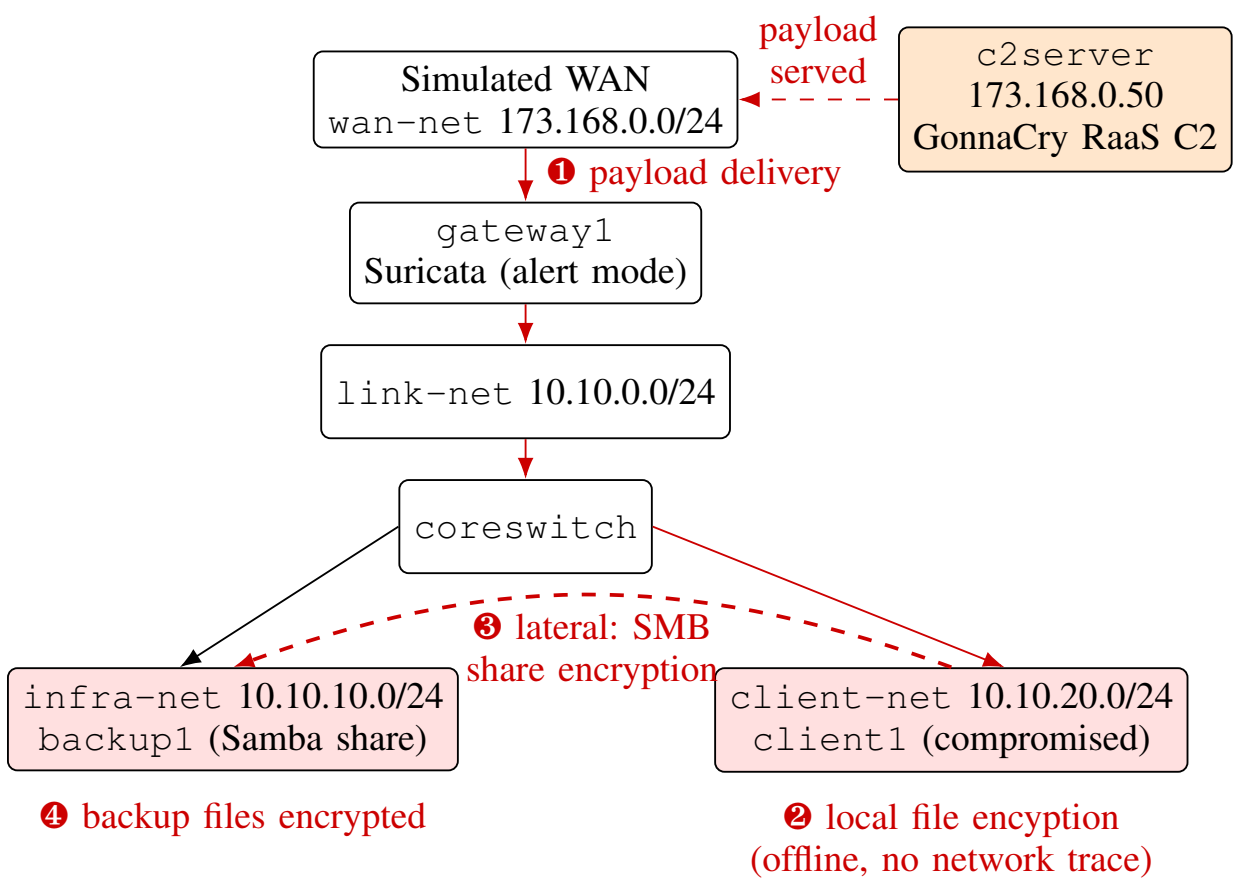
\begin{figure}[t]
    \centering
    \scalebox{0.88}{
    \begin{tikzpicture}[
        font=\footnotesize,
        node distance=5mm,
        box/.style={
            draw,
            rounded corners=2pt,
            align=center,
            inner sep=3pt,
            minimum height=7mm,
            fill=white
        },
        victim/.style={
            draw,
            rounded corners=2pt,
            align=center,
            inner sep=3pt,
            minimum height=7mm,
            fill=red!12
        },
        attacker_n/.style={
            draw,
            rounded corners=2pt,
            align=center,
            inner sep=3pt,
            minimum height=7mm,
            fill=orange!20
        },
        link/.style={
            -{Latex[length=1.6mm]},
            thin,
            black
        },
        graylink/.style={
            -{Latex[length=1.6mm]},
            thin,
            gray!50
        },
        atk/.style={
            -{Latex[length=1.6mm]},
            thin,
            red!80!black
        },
        lateral/.style={
            -{Latex[length=1.6mm]},
            thick,
            red!80!black,
            dashed
        }        
    ]
        \node[box] (wan) {Simulated WAN\\\texttt{wan-net} 173.168.0.0/24};
        \node[attacker_n, right=12mm of wan] (c2) {\texttt{c2server}\\173.168.0.50\\GonnaCry \gls{RaaS} C2};
        \node[box, below=4mm of wan] (gw) {\texttt{gateway1}\\Suricata (alert mode)};
        \node[box, below=3mm of gw] (link) {\texttt{link-net} 10.10.0.0/24};
        \node[box, below=3mm of link] (core) {\texttt{coreswitch}};
        \node[victim, below=7mm of core, xshift=-2.2cm] (infra) {\texttt{infra-net} 10.10.10.0/24\\ \texttt{backup1} (Samba share)};
        \node[victim, below=7mm of core, xshift=3.2cm] (client) {\texttt{client-net} 10.10.20.0/24\\ \texttt{client1} (compromised)};

        \draw[atk, dashed] (c2) -- node[above, font=\footnotesize, align=center] {payload\\served} (wan);
        \draw[atk] (wan) -- node[right, font=\footnotesize, xshift=1pt] {\ding{182} payload delivery} (gw);
        \draw[atk] (gw) -- (link);
        \draw[atk] (link) -- (core);
        \draw[link] (core.west) -- ([xshift=-4mm]infra.north);
        \draw[atk] (core.east) -- ([xshift=4mm]client.north);
        \draw[lateral, bend right=20] (client.north) to node[below, font=\footnotesize, sloped, align=center, text=red!80!black] {\ding{184} lateral: SMB\\share encryption} (infra.north);
        \node[font=\footnotesize, text=red!80!black, below=1mm of client, align=center] {\ding{183} local file encyption\\(offline, no network trace)};
        \node[font=\footnotesize, text=red!80!black, below=1mm of infra, align=center] {\ding{185} backup files encrypted};
    \end{tikzpicture}}
    \caption{Ransomware campaign. 
    }
    \label{fig:ransomware_attack}
\end{figure}

This attack demonstrates the ransomware campaign from payload delivery from an external \texttt{c2server}, client-side file encryption, lateral movement to the backup server, backup server's file encryption and key exfiltration back to the C2 server (as shown in Figure~\ref{fig:ransomware_attack}). The north--south path (client$\leftrightarrow$WAN) and the east--west path (client segment$\leftrightarrow$infra segment), produce logs for the labeled dataset.

The \texttt{c2server} runs on port~80 and implements the \gls{RaaS} backend. It serves two things: \texttt{/download-gonnacry/} delivers the encryptor binary, and \texttt{/decrypt/} accepts the victim's encrypted private key and returns the decrypted key after verifying payment. The server holds the RSA-2048 private key which is the root of the entire key hierarchy; without this, no victim can recover their files. The communications between the victim and the \texttt{c2server} are exchanged in plaintext on port~80. This is deliberately chosen for the testbed because the Suricata, deployed at the \texttt{gateway1} can easily perform deep packet inspection to match the rules on unencrypted packets, rather than dealing with additional complexity of analyzing encrypted payloads over HTTPS.

The ransomware attack is launched using GonnaCry~\cite{Gonnacry_Raas}, implemented as a Python~3 application, proceeds in the following stages.
\begin{enumerate}[nosep, label=(\roman*)]
    \item \textit{Payload delivery:} The compromised client sends a HTTP \texttt{GET} request to the \texttt{c2server} to download the ransomware binary. This communication gets captured at the \texttt{gateway1}, where Suricata \texttt{alert} rules log the connection, but do not block packets.
    \item \textit{Local file encryption:} The downloaded binary generates a fresh RSA-2048 key pair unique to this victim and encrypts targeted files, such as, documents, databases, source codes, VM disk images. Also, the extensions like \texttt{.pem}, \texttt{.key}, and \texttt{.p12} files are encrypted using randomly generated AES-256-GCM key. Each AES key is encrypted with the victim's RSA public key. The victim's RSA private key is then encrypted with the \texttt{c2server}'s public key and is stored locally at \texttt{\textasciitilde/gonnacry/}. Encrypted files are then renamed with \texttt{.GNNCRY} extension, and a \texttt{RANSOM\_NOTE.txt} containing machine ID and instructions for paying a ransom are written to the disk. This stage does not leave any trace at the gateway.
    \item \textit{Lateral movement and encryption:} GonnaCry also searches for backup services and mounts any SMB network shares that are reachable from the victim. Once identified, the ransomware encrypts backup files using the same AES-256-GCM/RSA-2048 scheme. This lateral encryption step destroys completely the backup archive as a recovery path, thereby forcing the victim to depend solely on the C2 server for decryption.
    \item \textit{Key exfiltration:} The victim sends the locally stored encrypted RSA private key along with the ransom to the \texttt{c2server}'s \texttt{/decrypt/} endpoint through HTTP \texttt{POST}. The server decrypts using its own RSA private key and return the victim's RSA private key to decrypt all AES keys and finally recover the files by the victim.
\end{enumerate}

\subsubsection{Layer-2 DHCP Starvation and Rogue Server}
\begin{figure}[t]
    \centering
    \scalebox{0.88}{
    \begin{tikzpicture}[
        font=\footnotesize,
        node distance=4mm,
        box/.style={
            draw,
            rounded corners=2pt,
            align=center,
            inner sep=2pt,
            minimum height=6mm,
            fill=white
        },
        victim/.style={
            draw,
            rounded corners=2pt,
            align=center,
            inner sep=2pt,
            minimum height=6mm,
            fill=red!12
        },
        attacker_n/.style={
            rounded corners=2pt,
            align=center,
            inner sep=2pt,
            minimum height=6mm,
            fill=orange!20
        },
        blind/.style={
            draw,
            rounded corners=2pt,
            align=center,
            inner sep=2pt,
            minimum height=6mm,
            fill=gray!15
        },
        link/.style={
            -{Latex[length=1.6mm]},
            thin,
            black
        },
        atk/.style={
            -{Latex[length=1.6mm]},
            thick,
            red!80!black
        },
        albl/.style={
            font=\footnotesize,
            text=red!80!black,
            align=center
        }
    ]

        \node[box] (wan) {Simulated WAN\\ \texttt{wan-net} 173.168.0.0/24};
        \node[blind, below=3mm of wan] (gw) {\texttt{gateway1}\\ Suricata (east--west blind spot)};
        \node[box, below=3mm of gw] (link) {\texttt{link-net} 10.10.0.0/24};
        \node[box, below=3mm of link] (core) {\texttt{coreswitch}\\DHCP relay};
        \node[victim, below=5mm of core, xshift=-2.2cm] (infra) {\texttt{infra-net} 10.10.10.0/24\\ \texttt{dhcp1} (legitimate DHCP)};
        \node[victim, align=center, inner sep=1pt, below=5mm of core, xshift=1.9cm] (client) {\texttt{client-net} 10.10.20.0/24};
        \node[attacker_n, below=2mm of client] (rogue) {Trudy (\texttt{client2})\\rogue DHCP server};

        \begin{scope}[on background layer]
            \node[draw, rounded corners=2pt, fill=red!12, inner sep=4pt, fit=(client)(rogue)](clientnet) {};
        \end{scope}

        \draw[link] (wan) -- (gw);
        \draw[link] (gw) -- (link);
        \draw[link] (link) -- (core);
        \draw[link] (core.east) -- (clientnet.north);

        \draw[atk] (rogue.west) to[out=180, in=-80] node[albl, left, pos=0.2, xshift=-6pt] {\ding{182} spoofed-MAC\\ \texttt{DISCOVER}} (core.south);

        \draw[atk] (core.west) -- node[albl, left, pos=0.3] {\ding{183} repeated \texttt{DORA}\\pool exhausted} (infra.north);

        \draw[atk] (rogue.east) to[out=70, in=-30] node[albl, right, pos=0.1, xshift=1pt] (lbl184) {\ding{184} rogue DHCP\\attacker\\gateway/DNS} ([xshift=20pt]client.south);

        \begin{scope}[on background layer]
            \node[draw, rounded corners=2pt, fill=red!12, inner sep=4pt, fit=(client)(rogue)(lbl184)](clientnet) {};
        \end{scope}

    \end{tikzpicture}}
    \caption{DHCP starvation and rogue DHCP server.
    }
    \label{fig:DHCP_attack}
\end{figure}
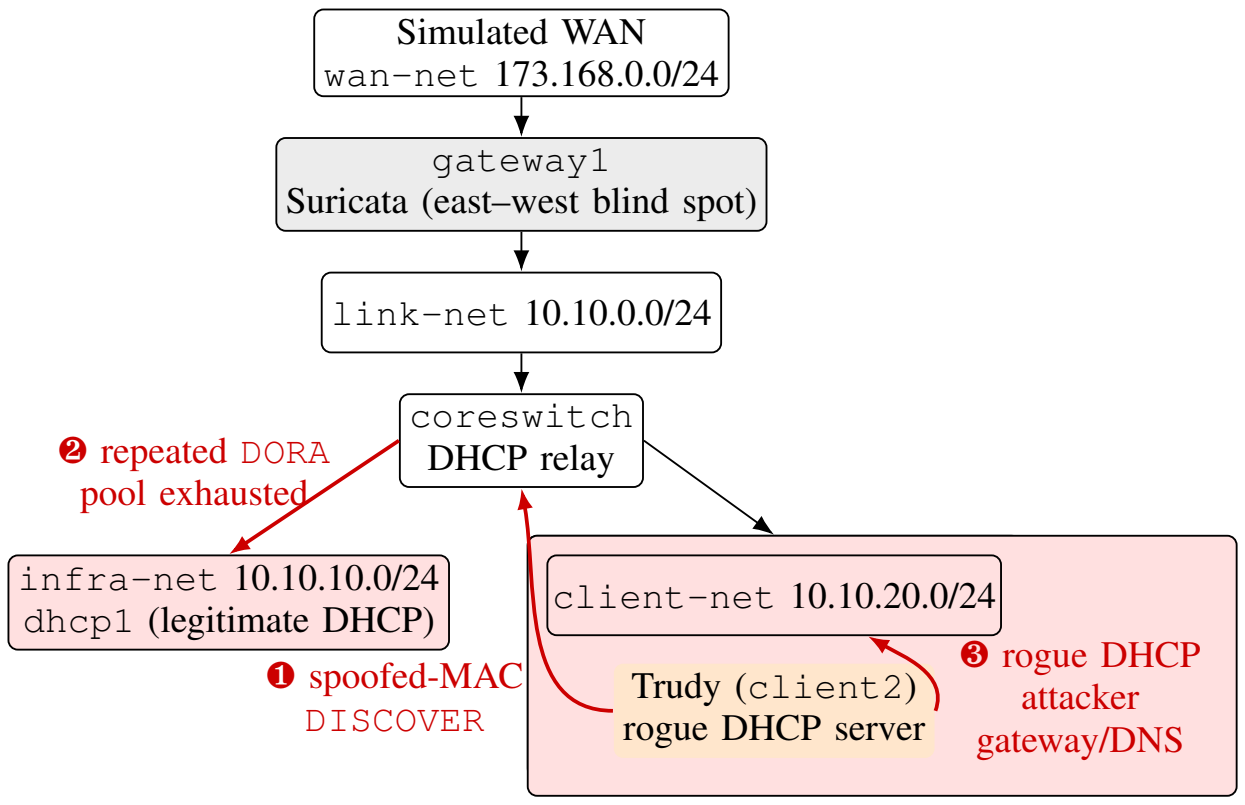

In this attack, Trudy is an insider operating from \texttt{client2} on the client segment who exhausts the legitimate DHCP IP address pool, thereby denying services to legitimate clients and coercing them to use a rogue DHCP server and its provided configurations. The attack, presented in Figure~\ref{fig:DHCP_attack}, runs in the following two coordinated phases from \texttt{client2}.
\begin{enumerate}[nosep, label=(\roman*)]
    \item \textit{DHCP pool exhaustion:} Trudy targets the legitimate DHCP server (\texttt{dhcp1}, 10.10.10.2) by sending a sequence of unique, fabricated DHCP \texttt{DISCOVER} messages via the \texttt{coreswitch}. As discussed in RFC~2131, the DHCP protocol does not have a built-in mechanism to authenticate clients and verify the legitimacy of the \texttt{DISCOVER} message. \texttt{dhcp1} responds to all these spoofed requests and commits a lease through the full \texttt{DORA} (\texttt{DISCOVER}--\texttt{OFFER}--\texttt{REQUEST}--\texttt{ACKNOWLEDGE}) handshake, until its address pool exhausts.
    \item \textit{Rogue DHCP server:} Trudy then instantiates a rogue server on \texttt{client2} with a small and different address range than the legitimate DHCP server. Now that the legitimate address pool is drained, new clients broadcasted \texttt{DISCOVER} messages are responded to by the rogue server. These new clients complete the \texttt{DORA} handshake with the rogue server and receive the configuration parameters like the default gateway, and the DNS resolver address. If Trudy wishes, then the default gateway can be set to \texttt{client2}'s own IP address, which allows Trudy to observe all the outbound traffic, which enables a transparent \texttt{MITM} attack. Otherwise, the rogue server responds with the real gateway address, so the connectivity to the clients appears normal.
\end{enumerate}

\subsubsection{Perimeter Web exploitation}
\begin{figure}[t]
    \centering
    \scalebox{0.88}{
    \begin{tikzpicture}[
        font=\footnotesize,
        node distance=5mm,
        box/.style={
            draw,
            rounded corners=2pt,
            align=center,
            inner sep=3pt,
            minimum height=7mm,
            fill=white
        },
        victim/.style={
            draw,
            rounded corners=2pt,
            align=center,
            inner sep=3pt,
            minimum height=7mm,
            fill=red!12
        },
        attacker_n/.style={
            draw,
            rounded corners=2pt,
            align=center,
            inner sep=3pt,
            minimum height=7mm,
            fill=orange!20
        },
        defense/.style={
            draw,
            rounded corners=2pt,
            align=center,
            inner sep=3pt,
            minimum height=7mm,
            fill=blue!10
        },
        safe/.style={
            draw,
            rounded corners=2pt,
            align=center,
            inner sep=3pt,
            minimum height=7mm,
            fill=green!10
        },
        blind/.style={
            draw,
            rounded corners=2pt,
            align=center,
            inner sep=2pt,
            minimum height=6mm,
            fill=gray!15
        },
        link/.style={
            -{Latex[length=1.6mm]},
            thin,
            black
        },
        atk/.style={
            -{Latex[length=1.6mm]},
            thick,
            red!80!black
        },
        blocked/.style={
            -{Latex[length=1.6mm]},
            thick,
            red!80!black,
            dashed
        }
    ]
        \node[attacker_n] (trudy){Trudy (external)\\on \texttt{wan-net}};
        \node[box, left=4mm of trudy] (wan) {Simulated WAN\\\texttt{wan-net} 173.168.0.0/24};
        \node[blind, below=4mm of wan] (gw){\texttt{gateway1}\\Suricata IPS + \gls{DNAT}:80};
        \node[box, below=4mm of gw] (link) {\texttt{link-net} 10.10.0.0/24};
        \node[blind, below=3mm of link] (core){\texttt{coreswitch}\\\texttt{iptables} inter-segment ACL};
        
        \node[victim, below=5mm of core, xshift=-2.5cm, minimum width=3.6cm] (infra) {
            \texttt{infra-net} 10.10.10.0/24\\\texttt{dns1}, \texttt{dhcp1}, \texttt{backup1}\\
            \textbf{\textcolor{red!80!black}{\texttt{web1} (compromised)}}
        };
        
        \node[safe, below=5mm of core, xshift=2.5cm] (client) {\texttt{client-net} 10.10.20.0/24\\\texttt{client1..N} (unreachable)};
 
        \draw[atk] (trudy) -- (wan);
        \draw[atk] (wan) -- node[right, font=\footnotesize, xshift=1pt] {HTTP request} (gw);
        \draw[atk] (gw) -- node[right, font=\footnotesize, xshift=1pt] {\gls{DNAT} forward} (link);
        \draw[atk] (link) -- (core);
        \draw[atk] (core) -- (infra);
        \draw[link] (core) -- (client);
 
        \draw[blocked] (infra.east) -- node[below, font=\footnotesize, text=red!80!black, align=center] {blocked\\by ACL} (client.west);
    \end{tikzpicture}}
    \caption{Trudy reaches \texttt{web1} from the simulated \gls{WAN} through the gateway's \gls{DNAT} rule (red solid path). The dashed line shows that the lateral movement from the infrastructure segment to the client segment is blocked by the \texttt{coreswitch} ACLs. Blue nodes indicate defensive controls, and the green node indicates the unreachable segment.}
    \label{fig:web_exploitation}
\end{figure}
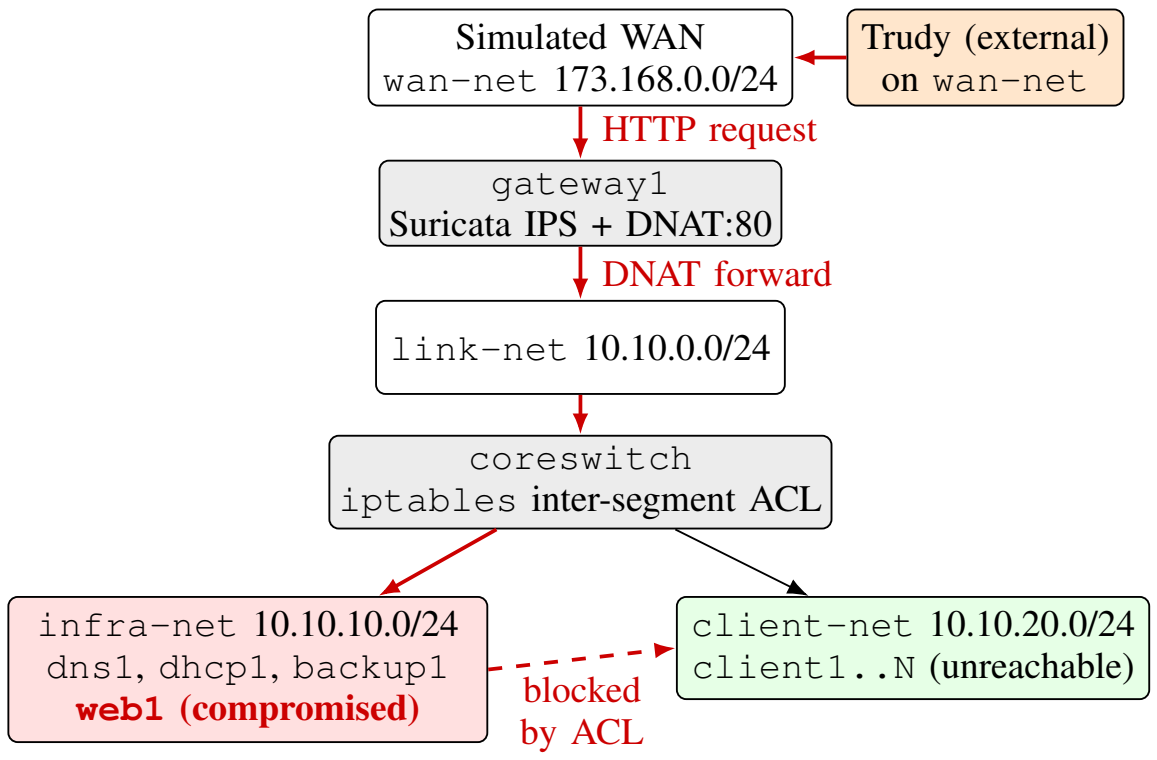
Figure~\ref{fig:web_exploitation} presents a web exploitation vector where Trudy exploits an Internet-facing web application vulnerability to gain access to the network, as an initial access step. This technique maps to the MITRE ATT\&CK ID T1190~\cite{MITRE_T1190}. The scenario considered is that an enterprise network hosts an internal web application, meant to serve internal subnets on plaintext HTTP; however, it can accidentally become externally reachable through a misconfigured firewall rule. Once exposed, the combination of the plaintext protocol and the application that was never hardened for public access creates an entry point for Trudy. The \gls{DNAT} rule on \texttt{gateway1} simulates an external exposure scenario, where HTTP traffic over TCP port~80 is forwarded to \texttt{web1}, rendering it to be reachable from the simulated WAN. Trudy, operating from the WAN, identifies this exposed service via network reconnaissance and initiates interaction. Another condition assumed in this scenario is that \texttt{web1} is deliberately misconfigured with two distinct vulnerabilities: a reflected SQL injection vulnerability and insecure login form that sends credentials in plaintext via HTTP \texttt{POST} requests. Using this, Trudy exploits the SQL injection vulnerability to exfiltrate sensitive data from the \texttt{web1} backend database, and is also in a position to intercept sensitive credentials transmitted insecurely through the form in plaintext.

\subsection{Defensive Measures}
\label{sec:Defensive_Measures}
This section provides defenses against each attack in Section~\ref{sec:Offensive_Measures}.
\subsubsection{Shared Control: Suricata Inline IPS at the Gateway}
All routed traffic through \texttt{gateway1} is forwarded to Suricata via \texttt{NFQUEUE} in inline prevention mode (dropping packets). The queue is configured as fail-open: if Suricata terminates or becomes saturated, traffic continues uninspected rather than causing a network outage. This design represents an explicit availability-over-security trade-off. The \texttt{HOME\_NET} variable encompasses the infrastructure, client subnets, and the gateway's WAN interface. The rule set combines community and local rules that detect activity, including ransomware communication between a compromised client and an attacker-controlled C2 server.

\subsubsection{Mitigating the Ransomware Campaign}
The compromised client contacts the \texttt{c2server} twice: once to download the payload and another to \texttt{POST} the encrypted private key for decryption. In both connections, the traffic flows through \texttt{gateway1}, where Suricata inspects them against rules matching \texttt{.onion} DNS resolution and domain generation algorithm (DGA) beaconing. Shifting the nature of these rules from \texttt{alert} to \texttt{drop} causes the payload download to be blocked before it even reaches the compromised client, which entirely prevents the infection. The following is the set of rules that mitigate the campaign.
\begin{lstlisting}[language=bash]
# Alert mode: logs but does not block
alert dns ... msg:"Tor .onion query"; sid:1000070;
# Drop mode: blocks payload
drop dns ...msg:"Tor .onion query"; sid:1000070;
\end{lstlisting}
We note that this detection relies on plaintext payload inspection, but the choice of using HTTP over HTTPS for C2 communication is discussed under the attacker model (refer Section~\ref{sec:Attacker_Model}). In the current configuration, Suricata operates in \texttt{alert} mode, which allows to log full ransomware lifecycle and collect the dataset. Even in the worst-case scenario where the payload gets downloaded on the client, the \textit{pull-based} backups keep the archive completely isolated. Because the backup server pulls the data rather than the client pushing it, the client has no way to reach or encrypt the archives during the attack. As an additional guideline, organizations can protect backup servers by storing backup encryption keys inside a Hardware Security Module (HSM)~\cite{HSM_NIST}.

\subsubsection{Mitigating the DHCP Layer-2 Attack}
The DHCP starvation attack entirely operates within the client segment's broadcast domain, which Suricata at the gateway cannot record. To mitigate IP pool exhaustion, we require a control to be placed at the bridge level, therefore we apply an aggregate rate limit at the client bridge, which restricts the number of DHCP \texttt{DISCOVER} messages 
forwarded per second.
\begin{lstlisting}[language=bash]
# Rate-limit DHCP DISCOVER on bridge
nft add rule bridge filter forward \
    udp dport 67 limit rate over 10/second \
    drop comment "dhcp-flood-guard"
\end{lstlisting}
This slows the exhaustion attack to the point where it becomes detectable before the IP pool is fully drained.

\subsubsection{Mitigating the Perimeter Web Exploitation}
The external entry point in ENTRAP is \texttt{web1}, which is reachable from the simulated \gls{WAN} through the gateway's \gls{DNAT} rule. To mitigate SQL injection and directory traversal attacks, Suricata rule set, deployed at \texttt{gateway}, matches these requests and drops them before they even reach the web server. In case, Trudy does compromise \texttt{web1}, network segmentation isolates it inside the \gls{DMZ}. However, security controls at \texttt{coreswitch1} enforces strict inter-segment ACLs, which restrict \texttt{web1} to initiate connection to the client segment or any other services (backup, DNS, DHCP) present within the infrastructure segment itself. To prevent lateral movement via credential reuse, all service-to-service communications mandate SSH key-based authentication.

\subsection{Setup and Experiment Cost}
\label{sec:Setup_and_Experiment_Cost}
\begin{table}[t]
    \centering
    \scriptsize
    \caption{ENTRAP build time from a clean state}
    \label{tab:setup_time}
    \begin{tabular}{@{}lr@{}}
        \toprule
        \textbf{Phase} & \textbf{Duration (s)} \\
        \midrule
        Network Creation (4 \texttt{libvirt} bridges) & 6 \\
        VM provisioning (7 entities + 2 clients, Cloud-init) & 227 \\
        Service Configuration (Ansible roles) & 719 \\
        Logging configuration & 103 \\
        \midrule
        \textbf{Total} & \textbf{1{,}055} \\
        \bottomrule
    \end{tabular}
\end{table}

\begin{table}[!b]
    \centering
    \scriptsize
    \caption{Payload delivery flows observed at \texttt{gateway1}}
    \label{tab:Ransomware_Logs}
    \begin{tabular}{@{}llrl@{}}
        \toprule
        \textbf{Phase} & \textbf{Direction} & \textbf{Size} & \textbf{Port} \\
        \midrule
        \multicolumn{4}{c}{\cellcolor{gray!30}\texttt{client}: 10.10.20.151 \quad \texttt{C2}: 173.168.0.50:80} \\
        \midrule
        Download 1 -- request  & \texttt{client} $\rightarrow$ \texttt{C2} & $\sim$580\,KB & :44406 \\
        Download 1 -- response & \texttt{C2} $\rightarrow$ \texttt{client} & $\sim$31.5\,MB & \\
        Download 2 -- request  & \texttt{client} $\rightarrow$ \texttt{C2} & $\sim$586\,KB & :56552 \\
        Download 2 -- response & \texttt{C2} $\rightarrow$ \texttt{client} & $\sim$31.9\,MB & \\
        \midrule
        \multicolumn{4}{c}{\cellcolor{gray!30}\textit{File encryption at compromised client
        -- no network trace}} \\
        \midrule
        Encrypted RSA key & \texttt{client} $\rightarrow$ \texttt{C2} & $\sim$5\,KB & :56554 \\
        Server\texttt{ACK} & \texttt{C2} $\rightarrow$ \texttt{client} & $\sim$2\,KB & \\
        \bottomrule
    \end{tabular}
\end{table}

Table~\ref{tab:setup_time} reports the time taken to build the entire enterprise system from a clean state on a single hypervisor host (Intel Xeon W-2255, CPU 10cores, 20 threads, 3.70GHz 32GB RAM, running Arch Linux). The setup proceeds in four stages: (1)~\textit{network creation}, which defines and starts the four isolated \texttt{libvirt} bridges; (2)~\textit{VM provisioning}, which creates each entity's disk image and Cloud-init ISOs, and boots all nine VMs; (3)~\textit{service configuration}, which uses Ansible over SSH to install and configure every service in order of dependency; and (4)~\textit{logging configuration}, which turns on protocol-split forensic logging across the environment. The biggest bottleneck is VM provisioning, mostly because of Cloud-init's first-boot initialization and package installation. A complete build from a clean state takes under 18 minutes. Because we use the same playbooks to reset the enterprise system, this time is also the absolute maximum time one has to wait to get to a clean state. Even if an experiment completely tears down the enterprise setup, it will be fully restored to the clean state within a single provisioning cycle.

Table~\ref{tab:Ransomware_Logs} shows the traffic recorded at \texttt{gateway1} between the compromised client ($10.10.20.151$) and the C2 server ($173.168.0.50$). In Transfer~1 and Transfer~2, the client requests the ransomware payload from the \texttt{c2server} and receives it to encrypt the files. The encryption runs 
locally on compromised client 
and hence leaves no trace at the gateway. Once the files are encrypted, the client submits the encrypted private key to the \texttt{c2server} and receives the decryption key to recover files.

\begin{table}[t]
    \centering
    \scriptsize
    \caption{DHCP starvation and rogue server attack}
    \label{tab:DHCP_Attack}
    \begin{tabular}{@{}lp{6.6cm}@{}}
        \toprule
        \textbf{Step} & \textbf{Observation and Effect} \\
        \midrule
        \rowcolor{gray!20} \multicolumn{2}{c}{\textit{Rogue DHCP server deployed}} \\
        Start daemon & \texttt{dnsmasq} starts on \texttt{client2} with IP pool .210--.220, and a 12-hour lease time.
        \\
        \midrule
        \rowcolor{gray!20} \multicolumn{2}{c}{\textit{Legitimate DHCP pool drained}} \\
        Lease flood & 49 forged MACs request IPs .152--.200 at $\sim$10 leases/s. Each lease lasts 7200 s, completely draining the real DHCP server's  IP pool. \\
        Forgery Proof & \texttt{hostname} and \texttt{client\_id} fields are empty on all 50 attack leases, whereas legitimate clients always include them. \\
        \midrule
        \rowcolor{gray!20} \multicolumn{2}{c}{\textit{Legitimate lease invalidation}} \\
        Renewal denied & Victim sends a \texttt{REQUEST} for .150, the rogue server replies with \texttt{NAK}, causing the victim to drop its current lease. \\
        Address seized & Released .150 is instantly grabbed by a fake MAC address, which prevents reclamation. \\
        \midrule
        \rowcolor{gray!20} \multicolumn{2}{c}{\textit{Victim bound to rogue server}} \\
        \texttt{DORA} on rogue & Full \texttt{DORA} runs through the rogue server. \texttt{client1} binds to .216, giving the attacker control over its gateway and DNS resolver. \\
        \bottomrule
    \end{tabular}
\end{table}
Table~\ref{tab:DHCP_Attack} shows a combined DHCP starvation and rogue server attack. The rogue server starts first, serving a range within the same subnet but distinct from the legitimate server's IP address pool. The legitimate DHCP server is then flooded with 49 consecutive forged requests, exhausting the address pool. The victim's renewal is refused by a rogue \texttt{DHCPNAK}, and its released address is seized, so its next \texttt{DORA} exchange can complete only with the rogue server. This gives Trudy control over the newly configured gateway and DNS resolver addresses of the affected client.

\section{Use Cases}
\label{sec:Use_Cases}
ENTRAP's clean state reset enables many use cases beyond the attacks and defenses mentioned above.
\begin{enumerate}[nosep]
    \item \textit{Dataset Collection:}
    We collect logs specific to each attack using a controller script that timestamps each run, executes the attack, and invokes an Ansible playbook to pull logs from every entity in parallel into a timestamped directory per run. These logs consist of network IDS events from the gateway, host telemetry (such as \texttt{auditd}, \texttt{syslog}, and \texttt{auth.log}) from the clients, and service logs (Kea DHCP, BIND9, Samba and Apache) from the infrastructure segment. Each run creates an attack-specific label directory to distinguish between different attacks.
    
    For the DHCP attack, the dataset contains Kea's lease table and the per-request DHCP \texttt{DORA} log, presenting the pool starvation under spoofed-MAC floods, alongside gateway traffic flow and anomaly events. For the ransomware campaign, the dataset captures compromised client-side \texttt{syslog}, gateway Suricata logs of C2 communication, and flow records, which provide the initial-access and encryption phase footprints. As a result of collecting known attack labels and precise timestamps, the dataset can be used to train supervised classifiers that distinguish between attack or benign traffic. And further samples from multiple clean state runs can serve as a validation dataset to evaluate the robustness of the model. We release our codebase, collected logs (Suricata JSON files, \texttt{auditd} traces, Kea DHCP logs, and \texttt{pcap} files) and labeled datasets publicly on our repository~\cite{ENTRAP_GitHub}.
    
    \item \textit{Replicating \& Hardening Custom Enterprise} 
    \textit{Environment:}
    Because the testbed is defined in Ansible playbooks, 
    cybersecurity researchers 
    can modify configurations to reflect their own enterprise system. Updating subnets, adding services, or changing policies requires only editing the relevant playbook. Once deployed, users can launch attacks, inspect each entity, and evaluate defensive measures. If an entity breaks during an attack, then it can be destroyed and rebuilt to a clean state. This cycle of configuring, attacking, evaluating, and resetting provides a safe way to evaluate defenses before applying them in production enterprise environments.
    
    \item \textit{Cryptographic Analysis of Ransomware Encryption Schemes:}
    The ransomware campaign uses AES-256-GCM, implemented as a module that can be substituted with a different algorithm. Because the testbed can be easily reset to a clean state, a cryptanalyst can substitute a different algorithm on each run to collect the resulting plaintext and ciphertext pairs. Across multiple runs, researchers can study their susceptibility to chosen-plaintext attacks.
    
    \item \textit{Learning Enterprise Network Assembly:}
    ENTRAP provides hands-on experience in building a realistic enterprise network in the correct order. The Ansible playbooks map out entity roles and their dependencies explicitly, so
    students and practitioners
    can read the code and understand exactly why one entity must be configured before the next. This ordered assembly replicates what an administrator faces in a real enterprise, but in a safe environment where mistakes carry no consequences.
\end{enumerate}

\section{Ransomware Detection Experiment}
\label{sec:Ransomware_Detection_Experiment}
We explore the \textit{dataset collection} use case described in Section~\ref{sec:Use_Cases}. We train a Random Forest (RF) classifier on our testbed logs and evaluate its robustness on another run on the same testbed. We executed three GonnaCry ransomware runs and one baseline run on the testbed, each starting from a clean state snapshot, producing 9,068 labeled events (5,282 benign and 3,786 ransomware). Events falling within the recorded attack window are labeled \texttt{gonnacry}; and those outside (normal system activity on hosts, and the idle attacker VM) as \texttt{benign}. These samples combine Suricata flow records from \texttt{gateway1} with \texttt{auditd} syscall traces from the client VMs. Rather than using raw numeric fields, we have 24~behavioral features from the logs: seven network-level features from Suricata (\texttt{bytes\_toserver}, \texttt{bytes\_toclient}, \texttt{pkts\_toserver}, \texttt{pkts\_toclient}, \texttt{flow\_duration}, \texttt{dest\_port}, \texttt{src\_port}), binary flags from the ransomware executable (\texttt{exe\_is\_gonnacry}, \texttt{exe\_from\_tmp}), one-hot encoded syscall types (\texttt{rename}, \texttt{execve}, \texttt{connect}, \texttt{openat}, \texttt{unlink}), audit key categories (\texttt{file\_rename}, \texttt{net\_connect}, \texttt{tmp\_write}, \texttt{exec}), encryption path markers (\texttt{.GNNCRY}, \texttt{gonnacry/}), and destination flags for the C2 and backup servers. The training samples were split 80:20 with stratified sampling (7,254:1,814). To evaluate robustness, we re-ran a single ransomware attack on a clean state enterprise network setup and collected 3,573 new samples (1,685 benign, and 1,888 ransomware) that the model had never encountered. Table~\ref{tab:ML_Results} reports the results.

\begin{table}[t]
    \centering
    \caption{Ransomware detection}
    \label{tab:ML_Results}
    \begin{tabular}{@{}lcccc@{}}
        \toprule
        \textbf{Evaluation} & \textit{Accuracy} & \textit{Precision} & \textit{Recall} & \textit{F1-score} \\
        \midrule
        Trained (80:20) & 0.92 & 1.00 & 0.80 & 0.89 \\
        Inferred & 0.89 & 1.00 & 0.79 & 0.88 \\
        \bottomrule
    \end{tabular}
\end{table}

\vspace{0.2in}
On the trained samples, the classifier achieves 92\% accuracy with a precision of 100\% at a recall of 80\%. On the evaluation dataset, accuracy drops marginally to 89\% at the same precision of 100\% (zero false positives), at a recall of 79\%. The minimal drop in recall confirms that the behavioral features generalize across clean-state rebuilds.

\section{Conclusions and Future work}
\label{sec:Conclusion}
We developed \gls{ENTRAP}, an \gls{IaC} enterprise cyber range that defines a complete segmented enterprise network as version-controlled automation code and rebuilds the environment identically from a clean state on demand. We launched a ransomware campaign, DHCP starvation and a rogue DHCP server, and perimeter web exploitation attacks and deployed corresponding defensive measures. As each attack and log stream is defined as code and each experiment starts from the clean state, ENTRAP can be used to generate reproducible, labeled datasets for these attacks. Using the collected dataset, we trained a Random Forest classifier for ransomware detection and demonstrated that the behavioral features generalize across independent clean state runs, achieving 89\% accuracy and zero false positives. We release the ENTRAP codebase, logs, and labeled datasets in this GitHub repository~\cite{ENTRAP_GitHub}.

%
\vspace{0.2in}
\textbf{Lessons Learned and Research Implications:} We observed several insights working with ENTRAP. First, the pull-based backup approach survived the ransomware campaign because it pulls files from the clients instead of letting the clients push files to the backup server. This simple choice eliminates lateral movement and the possibility of encrypting backup files from a compromised client. Second, we noticed that the DHCP starvation attack runs entirely within the client segment's broadcast domain, and Suricata IPS at the gateway cannot observe it. This shows that the enterprise network needs east--west monitoring at the bridge level, not just north-south inspection at the gateway. Third, even if a web server is compromised, the coreswitch ACLs prevents lateral movement to the client segment, showing that dividing the network into separate segments limits the blast radius of a perimeter breach.

As a part of future work, (i)~we plan to integrate Ansible AWX as a web-based interface, enabling authenticated users to provision, reset, and interact with the testbed through a browser with a role-based access control for multi-user learning, (ii)~we plan to include a VPN gateway to model a perimeter-less enterprise network, and extend Suricata's detection to identify HTTPS-based C2 communication using SNI-based rules and TTLS fingerprinting, and  (iii)~we wish to build robust detection models using the datasets collected from our testbed and evaluate them against zero-day attacks.

\balance

\begingroup
\newpage
\bibliographystyle{IEEEtran}
\bibliography{references}
\endgroup

\end{document}